\documentclass{aa}  
\usepackage{graphicx}
\usepackage{natbib}
\usepackage{txfonts}
\begin{document}
   \title{Spatially resolving the hot CO around the young Be star
     51~Ophiuchi\thanks{Based on ESO/VLTI programs
       077.C-0706 and 60.A-9054}}

   %\\subtitle{}

   \author{E. Tatulli
          \inst{1,2}
          \and
          F. Malbet \inst{2}	  
\and
	  F. M\'enard \inst{2}
          \and
          C. Gil \inst{3}
	  \and
	  L. Testi \inst{1,4}
	  \and
	  A. Natta \inst{1}
          \and 
          S. Kraus \inst{5}
          \and
          P. Stee \inst{6}
          \and
          S. Robbe-Dubois \inst{7}
          }

   \offprints{Eric.Tatulli@obs.ujf-grenoble.fr}

   \institute{INAF-Osservatorio Astrofisico di Arcetri, Istituto Nazionale di
  Astrofisica, Largo E. Fermi 5, I-50125 Firenze, Italy
         \and
             Laboratoire d'Astrophysique de Grenoble, UMR 5571
  Universit\'e Joseph Fourier/CNRS, BP 53, F-38041 Grenoble Cedex 9, France
	     \and
             European Southern Observatory, Casilla 19001, Santiago 19,
  Chile
             \and
           ESO, Karl-Schwarzschild Strasse 2, D-85748 Garching bei
           Muenchen, Germany
           \and
           Max Planck Institut f\"ur Radioastronomie, Auf dem H\"ugel 69, 53121 Bonn, Germany
           \and
           UMR 6525 CNRS H. FIZEAU – UNS, OCA, Campus Valrose, F-06108
           Nice cedex 2, France,  CNRS - Avenue Copernic, Grasse,
           France. 
           \and
           Laboratoire Universitaire d'Astrophysique de Nice, UMR 6525
  Universit\'e de Nice/CNRS, Parc Valrose, F-06108 Nice cedex 2, France
         }

   \date{Received ; accepted }

% \abstract{}{}{}{}{} 
% 5 {} token are mandatory
 
  \abstract
  % context heading (optional)
  % {} leave it empty if necessary  
   {}
  % aims heading (mandatory)
   {51 Oph is one of the few young Be stars displaying 
a strong CO overtone emission at $2.3$ microns in addition to the
  near infrared excess commonly observed in this type of stars. In
  this paper we first aim to locate the CO bandheads emitting region. Then,
  we compare its position with respect to the region emitting the near infrared continuum.}
  % methods heading (mandatory)
   {We have observed 51 Oph  with AMBER in
     low spectral resolution (R=35), and in medium spectral resolution
     (R=1500) centered on the CO bandheads. } 
  % results heading (mandatory)
   {The medium resolution AMBER observations clearly resolve the CO
     bandheads. Both the CO bandheads and continuum emissions are
     spatially resolved by the interferometer. Using simple
     analytical ring models to interpret the measured visibilities,
     we find that  the CO bandheads emission region is 
     compact, located at $0.15_{-0.04}^{0.07}$AU from the star, and
     that the adjacent continuum is coming from a region further away ($0.25_{-0.03}^{0.06}$AU).  These results confirm
     the commonly invoked scenario in which the CO bandheads originate
     in a dust free hot gaseous disk. Furthermore, the continuum
   emitting region is closer to the star than the dust sublimation
   radius (by at least a factor two) and we suggest that hot gas
   inside the dust sublimation radius significantly contributes to the
   observed 2 $\mu$m continuum emission.}
% We
%   have obtained a spectrum in which CO bandheads are detected and
%   resolved, as well as visibilities as a function of the wavelength. 
%   We find that 51 Oph is spatially resolved both in the continuum and
%   in the CO lines. \textbf{We directly locate the CO bandheads emission region at
%     a distance  of $0.15_{-0.04}^{0.07}$ AU from the star. We show,
%     independently of any modelling, that this region
%     is smaller than that of the near infrared continuum which inner
%     rim is located at $0.25_{-0.03}^{0.06}$AU. } This result is
%   consistent with the picture unveiled by spectroscopic modelling in
%   which the CO is originating from a \textbf{dust free} hot gaseous disk
%   in rotation around the star. \textbf{However, the measured continuum emission
%     region being at least twice smaller than the dust sublimation
%     radius (assuming that the dust is in equilibrium with the
%   radiation field), we} further suggests
%   that the hot gas also significantly contributes to the infrared continuum emission.}
  % conclusions heading (optional), leave it empty if necessary 
  % { }
   {}


   \maketitle
%
%________________________________________________________________

\section{Introduction}
The environment of young stars is formed by a dusty and gaseous
circumstellar disk in which the first stages of planet formation are
thought to take place. Hence, building a comprehensive picture of the physical
mechanisms governing this formation requires the understanding of the
distribution and the evolution of both the dust and gas species.
The content and dynamics 
of the gaseous component, which dominates the total mass
of the disk, control the accretion-ejection processes and will 
define the final architecture of forming planetary systems. 
On the other hand, the dust, responsible for the continuum infrared
excess, plays an essential role in determining the
structure of the protoplanetary disk itself,
since its opacity dominates in the continuum.  
Dust grains are also the first building blocks
that will grow into kilometric planetesimals and then into eventual
planets. \\
To study the planet formation taking place in the very inner region of
the disk, one needs to combine milli-arcsecond high {\it angular}
resolution techniques with high {\it spectral} resolution in order
to {\it directly} probe the physical processes at stake both in the
continuum and in the emission lines. 
In this respect, near infrared spectro-interferometry that associates 
long baseline near infrared interferometry with spectroscopy appears
perfectly suited. This technique indeed offers a unique way to
  spatially resolve both the continuum infrared excess and emission line components,
  that is to locate the regions of emission independently of any
  prior assumption regarding the physical mechanisms at play. 
In that sense, infrared spectro-interferometry has already provided important results
about the origin of the hydrogen line $\mathrm{Br}\gamma$. 
Different authors have thus located the $\mathrm{Br}\gamma$ emission region, showing that it can arise either from
the accreting columns of gas falling onto the star \citep{eisner_1} or from
outflowing winds in which ionized matter is ejected from the star
\citep{malbet_1, tatulli_2, kraus_1}. \\
In this paper we are interested in exploring lines which are more
direct probes of gas in the disk itself.  
We present the first interferometric observations of the
$2.3\mu\mathrm{m}$ CO overtone emission in the (B9) Be star 51 Oph.  
51 Oph is one of the very few young stars where this emission is strong
enough to be observed with infrared
spectro-interferometry \citep{thi_1,berthoud_1}. This star exhibits
also a strong $\mathrm{Br}\gamma$ emission, which, assuming a
magnetospheric origin \citep{muzerolle_1}, indicates an accretion
rate of $\dot{M}=1-2.10^{-7}M_{\odot}/yr$ \citep{garcialopez_1,
  brittain_1}. The dust component  is rather compact (\citet{lagage_1,
  liu_1}) and shows a strong silicate
feature in emission \citep{bouwman_1} but no PAH \citep{meeus_1}, consistent
with a flat and geometrically thin dusty disk \citep{malfait_1}. All the
previous analysis are pointing towards a disk seen nearly edge-on.
%__________________________________________________________________
\section{Observations and data reduction}
 \begin{table*}[!t]
\caption{\label{tab_log}Log of the observations and computed absolute
  visibilities in the continuum.}
\begin{tabular}{cccccccc}
\hline 
Date & Telescopes & Baseline (m)  & PA ($^{\circ}$)& Wavelength ($\mu$m) & Resolution & Calibrator &
Absolute visibility  \\   \hline
10-09-2006 & U1-U2-U4 & 55/82/121 &
34/91/69 & 2.3 & 1500 & HD170499 &
1.0$\pm$0.1/0.8$\pm$0.05/0.8$\pm$0.03\\
05-09-2007 & G1-D0-H0 & 69/60/71 &
326/80/17 & 2.2 & 35 &  HD163955  &
0.8$\pm$0.1/0.85$\pm$0.1/1.0$\pm$0.1\\
05-09-2007 & G1-D0-H0 & 65/46/70 &
345/89/24 & 2.2 & 35 &  HD163955  &
1.0$\pm$0.15/1.0$\pm$0.1/0.87$\pm$0.1\\
06-09-2007 & G1-D0-H0 & 64/31/67 &
0/98/27 & 2.2 & 35 &   HD172051  &
0.85$\pm$0.12/0.95$\pm$0.1/0.95$\pm$0.1\\
\hline 
\end{tabular}
\end{table*}
 \begin{figure}[!t]
   \centering
   \includegraphics[width=0.3\textwidth]{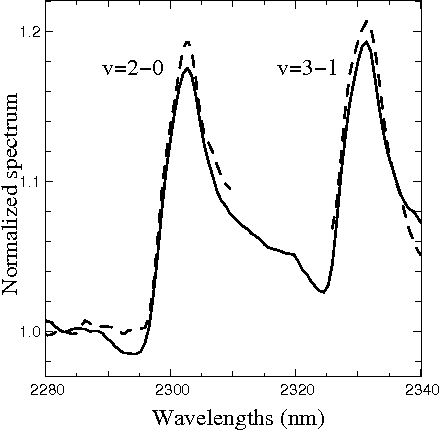}
   \caption{\label{fig_data}AMBER calibrated spectrum of 51 Oph
   around the 2-0 and 3-1 bands of the CO overtone at $2.3$microns. For comparison purposes is
   plotted (dashed line) the same spectrum measured with the TNG spectrograph
   (L. Testi, private communication). Note that we did not plot the
   TNG spectrum between the two bandheads  because of irrelevant instrumental artifact.}%
   \end{figure}
51 Oph ($K=4.3$) has been observed with AMBER, first in medium spectral
resolution (MR-K, R=1500) with the 8m Units Telescopes (UTs) of the VLTI in September
2006, then in low resolution  (LR-K, R=35) with the 1.8m Auxiliary
Telescopes (ATs) in September 2007. The log of the observations is presented in Table \ref{tab_log}.\\ 
{\it Spectrally dispersed observations:} The observations in medium
spectral resolution have been performed in the wavelength range
$[2280, 2340]\mu\mathrm{m}$ around the CO overtone bandheads emission.  
On the AMBER detector are recorded both
the dispersed photometry from which the spectrum of our
target is derived, and the dispersed interference pattern from which the visibilities as a function of the wavelength are computed.
A reference source is required to calibrate both observables (i.e.  the
spectrum and the visibilities). \\
{\it 51 Oph calibrated spectrum:} In order to correct for the wavelength dependent telluric and instrumental features in the AMBER
  spectra, we used the observations of the calibration star
  HD170499. This star has a  K4III spectral type and has intrinsic
  photospheric absorption features that need to be removed before using its observations to derive the response of the atmosphere and instrument. In
  particular, cool stellar photospheres show strong CO absorption
  longward of 2.9~$\mu$m (e.g. \citet{wallace_1}). To correct for the
  photospheric emission we constructed a template photospheric spectrum averaging three of the best quality spectra of stars with spectral type as
  close as possible to our own from \citet{wallace_1}: HR~1457
  (K5III), HR~3275 (K4.5III), and HR~6705 (K5III). The average
  spectrum (normalized to the continuum) was then multiplied by the spectral slope of a K4 III star using the \citet{pickles_1} models available on the ISAAC
  web pages at ESO to obtain the template. The template spectrum was
  then smoothed to the AMBER spectral resolution and used
  to remove the photospheric signatures in the calibrator spectrum and
  derive the response of the system. This response was then used to
  correct the spectrum of 51~Oph. In Fig. \ref{fig_data}
  we show the final AMBER spectrum compared to a
  K band spectrum of 51~Oph obtained at the TNG telescope with the
  NICS NIR spectrograph (Isella, priv. comm.). % To correctly analyze the
% spectral behavior of the visibility in the line, one needs to correct the AMBER
% spectrum from wavelength dependent instrumental effects, for example by dividing it by the spectrum of the calibrator (corrected from its
% spectral slope). Unfortunately, in our particular case, the 
% calibrator displays photospheric CO absorption lines. In order to
% remove these features which are originating from the reference star and
% to obtain the true instrumental response of the instrument, we 
% proceeded as follow:  (i) we took the template spectrum of K4III
% star from \citet{wallace_1} smoothed to the AMBER spectral resolution, then
% (ii) we corrected it from the appropriate spectral slope fitting a
% polynomial to the low resolution spectrum data given in the ISAAC
% archive, and finally (iii) we divided the observed spectrum of
% the calibrator by the template.  The calibrated spectrum of 51 Oph is
% then obtained by dividing the AMBER one by the instrumental
% response, computed as described above. Finally we normalized the
% spectrum to unity in the continuum, that is for the  $[2280,
% 2300]\mu\mathrm{m}$  spectral range. The results in shown in Fig \ref{fig_data} and is in excellent agreement with the
% spectrum measured with the TNG spectrograph. {\bf method to be precised by
%   Leonardo + short description of TNG data}
\\
{\it 51 Oph dispersed calibrated visibilities:} We computed
visibilities using the standard data reduction algorithms described in
\citet{tatulli_1}, previously applying the AMDC sofware
\citep{licausi_1} to take away the spurious
fringes of the detector that can bias the estimation of the
visibility. A drastic fringe signal-to-noise ratio (SNR) selection
using the $20\%$ best frames was performed (for both the source
and the calibrator)  to insure reliable absolute calibrated
visibilities \citep{tatulli_1}. These latter were obtained by
dividing the 51 Oph raw visibilities by the reference ones, assuming
a diameter of $1.28$mas for the calibrator, according to the catalog of
\citet{merand_1}. Finally, we binned the spectrally
dispersed visibilities in the wavelength direction in order to
increase the SNR. The results are presented in
Fig. \ref{fig_results_MRK_alone} and the values of the visibilities in
the continuum (i.e. average on the continuum spectral range) are given in Table \ref{tab_log}.  \\
{\it 51 Oph low resolution visibilities:}  LR-K data were reduced
following the same procedure as the MR-K ones and then 
averaged spectrally to derive one single visibility of the continuum
per baseline, whose values are plotted in
Fig. \ref{fig_results_LRK_alone} and given in Table \ref{tab_log} as well. 
%__________________________________________________________________
\section{Modelling}
\begin{figure}[!t]
   \centering
   \includegraphics[width=0.34\textwidth]{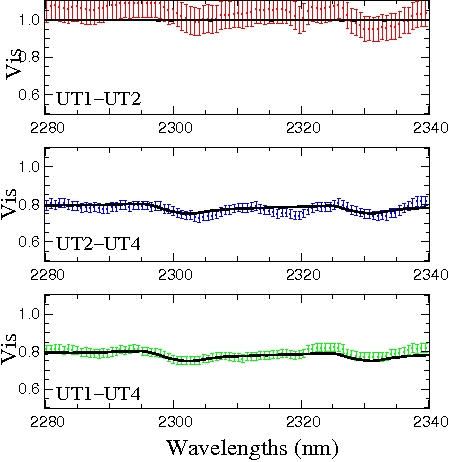}
   \caption{MR-K calibrated visibilities as a function of the
   wavelengths for the three  baselines and
   overplotted best model (black thick line), for the ring case. 
   % Bottom: LR-K calibrated visibilities as a function of the
%    projected baselines, for the three observations, and overplotted
%    best model (lines). In each panel,
%    colors red, blue,and green correspond to baselines G1-D0, D0-H0,
%    and G1-D0 respectively.
 }\label{fig_results_MRK_alone}%
   \end{figure}

\begin{figure}[!t]
   \centering
   \includegraphics[width=0.32\textwidth]{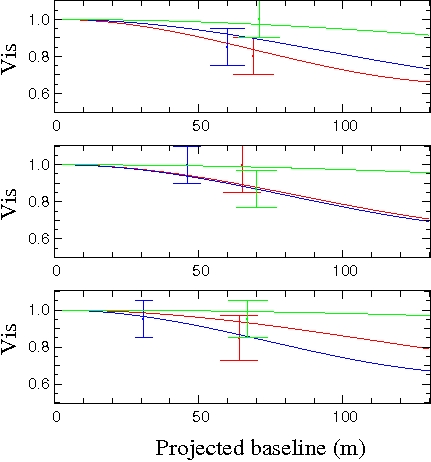}
   \caption{LR-K calibrated visibilities as a function of the
   projected baselines, for the three observations, and overplotted
   best model (lines). In each panel,
   colors red, blue,and green correspond to baselines G1-D0, D0-H0,
   and G1-D0 respectively.}\label{fig_results_LRK_alone}%
   \end{figure}

Our MR-K data resolve spectrally the two CO bands
(although not the rotational components within each band). 
For the longest baselines UT2-UT4 and UT1-UT4, the emission is
spatially resolved at all wavelengths, though with rather high
visibilities of 0.8. Our LR-K data are consistent with the MR-K ones,
displaying high visibilities between 0.8 and 1 in the continuum at 2.2$\mu$m.\\
To derive more information, we need to compare the observed
visibilities to model predictions. We adopt simple geometrical models
and describe both the continuum and bandheads emitting regions as coming
from rings of constant surface brightness. At each wavelength we
assume that the emission is the sum of the unresolved stellar flux
$F_s$, of a continuum excess emission $F_c$ and of emission in the CO
bandheads $F_l$, with ratios that we derive from observations. In
particular, the ratio $F_c/(F_s+F_c)$ is derived from the spectral
energy distribution (SED) photometric data \citep{waters_1}, fitting
a Kurucz spectrum  \citep{kurucz_1} for the star with $T_{\ast}=10000$K,
$R_{\ast}=5.3R_{\odot}$. We find $F_c/(F_s+F_c) = 25\%$ and $F_c/(F_s+F_c) = 30\%$ at 2.2$\mu$m and  2.3$\mu$m, respectively. The
ratio $F_l/(F_s+F_c+F_l)$ is given by our normalized AMBER spectrum
(see Fig. \ref{fig_data}, left) in the CO bandheads and is about $18\%$
at the peak of the 2-0 band and $19\%$ at the peak of the 3-1 band.\\
We compute two families of models, one with narrow\footnote{the width of the ring is fixed at the
  typical value of $\Delta{R}/R=20\%$, the results being
  poorly dependent on its value as long as
  $\Delta{R} \ll R$.} rings and one with broad rings
extending from the stellar radius (in the following, uniform brightness). 
The parameters of the model are the inclination ($i$), the position
angle ($PA$) of the system and the size of the bandheads ($R_l$) and
continuum ($R_c$) emitting region, respectively. The visibility as a function of the wavelength is given by: 
\begin{equation}
V^{\lambda} =\frac{F_s^{\lambda} + F_c^{\lambda}V_c^{\lambda}(R_c, i ,
  PA) + F_l^{\lambda}V_l^{\lambda}(R_l, i ,
  PA)}{F_s^{\lambda}+F_c^{\lambda}+F_l^{\lambda}}
\end{equation}
where $V_c^{\lambda}$ is the visibility of the region emitting the
continuum and $V_l^{\lambda}$ that of the region emitting the bandheads.\\
Fitting together the continuum and bandheads models, we have
performed a least-square minimization test on a large grid of
parameters varying both radii from 0 to 3AU, the inclination from
0 to 90$^{\circ}$ and the position angle from 0 to 180$^{\circ}$.
The $\chi^2$ function presents a clear and unambiguous minimum, for two reasons: (i) the high value of the visibilities imposing that we
are in the first lobe of the visibility function, hence that the
emission regions  are rather compact and, (ii) the broad range of
baseline orientations which put strong constraints on the inclination
and the position angle of the system.  
% More precisely we note that, as
% long as we are investigating the first lobe of the visibility of the
% models, the LR-K data are hardly adding constraints on the
% model-fitting, given that their error bars are 5 to 10 times bigger than the
% MR-K ones. However, these LR-K measurements are
% allowing us to definitely rule out a second potential
% solution for the diameter of the continuum emission region, which
% would have been roughly twice bigger, corresponding to a fit in the
% second lobe of the model visibility.
The parameters of the best-fit model 
are summarized in Table \ref{tab_res}, and the
result of the fit is shown in Fig. \ref{fig_results_MRK_alone} and \ref{fig_results_LRK_alone}, for the
narrow ring model case.\\
We note that both the bandheads and continuum components are unresolved for
the smallest UT baseline shown in the top panel of
Fig. \ref{fig_results_MRK_alone}. This suggests that the small variations of the visibility detected in the CO bandheads 
 are probably remaining biases of the data reduction, given the low SNR of
the interferograms.  However, considering the error bars
of the measurements, this (possible) bias has only a  very marginal influence on the sizes we derive and does not impact on their interpretation. 
% Examples of 2D-$\chi^2$ maps are given in Fig. \ref{fig_results_chi2} in
% the case of the narrow ring model, and 
%   \begin{figure}[!t]
%    \centering
%    \includegraphics[width=0.45\textwidth]{/home/etatulli/WORK/AMBER/51 Oph/51oph_results_chi2.jpg}
%    %%%\includegraphics{empty.eps}
%    %%%\includegraphics{empty.eps}
%    \caption{$\chi^2$ maps of the ouput parameters of our
%    the ring model. Contours are plotted starting at 
%    1$\sigma$, with 1$\sigma$ spacing. \label{fig_results_chi2}}
%    \end{figure}
\begin{table}[!t]
\caption{\label{tab_res}Best-fit model parameters.}
\begin{tabular}{cccccc}
\hline
Model & $\chi^2_{min}$ & $i$ ($^{\circ}$)& $PA$ ($^{\circ}$)&
$R_c$ (AU)& $R_l$ (AU)\\
\hline
Uniform Brightness & 0.8 & $85_{-15}^{+5}$& $129_{-9}^{+10}$& $0.43_{-0.15}^{+0.10}$& $0.36_{-0.07}^{+0.06}$\\
Narrow Ring & 0.8 & $82_{-15}^{+8}$& $126_{-5}^{+15}$&
$0.24_{-0.03}^{+0.06}$ & $0.15_{-0.04}^{0.07}$\\
\hline
\end{tabular}
\end{table}
%__________________________________________________________________

\section{Discussion}
\subsection{Sizes of emitting regions vs. modelling}
The uniform brightness and the narrow ring are the two
extreme cases of a unique model of uniform distribution in which the width of
the emitting region is varying. We emphasize that this
  geometrical modelling defines the metric that allows us to unequivocally link the measured
  visibilities to the spatial brightness distribution of these regions,
  through the Zernicke-Van Cittert theorem
  \citep{goodman_1}. It does not introduce assumptions
  regarding the mechanisms at the origin of these emissions. 
We present in Table \ref{tab_res} 
the results for the two scenarios. All the intermediate models
would also work, with intermediate values for the sizes. Note that 
the uniform brightness returns the largest size because emission is added
at small spatial scales down to the stellar radius. The
uniform brightness  has no proper physical meaning since neither the
dust nor the CO gas can survive so close to the star. However, it
represents the {\it characteristic size} of the emission region. As this
quantity is commonly used in infrared interferometry, it is given
here for this purpose. 
Conversely the narrow ring model is sensitive to the
location where most of the emission is arising from, usually the {\it inner
  rim} of this emission. In the following, we will analyze our results
in the light of the narrow ring only. We emphasize that the size of
the CO emitting region is {\it smaller than the one of the continuum},
independently of the choice of our model. 
\subsection{Inclination and position angle of the system} 
We find a position angle of
$126_{-5}^{+15}$ degrees and an inclination of $i=82_{-15}^{+8}$ degrees. This
inclination is in agreement with most of the previous studies. 
\citet{berthoud_1} have, however, proposed an
alternative scenario in which the object would have an inclination
lower than $36^{\circ}$. Our measurements are positively ruling out
this latter option.
\subsection{Locating the CO bandheads emission region} 
For the first time, we are
spatially resolving the CO bandheads emission region in
51 Oph. We find that most of the CO bandheads are emitted at a radius of
$R_l =0.15_{-0.04}^{0.07}$AU from the central star. This is fully supporting 
the scenario in which  the CO bandheads emission is arising from the
inner region of a hot gaseous disk in Keplerian rotation, between  $0.15$AU and
$0.35-0.53$AU, as derived from spectroscopic modelling of the CO
bandheads \citep{thi_1, berthoud_1}, and where the
  temperature is hot enough ($\ge 2000$K) to excite the first overtone
  bands at 2.3$\mu$m. More precisely, these authors, though they
are fitting the CO bandheads with a broad ring, are pointing out that
most of the emission comes from a region close to the inner radius of
their gaseous disk model at $\sim$0.15AU. This is in excellent
agreement with the value we derive in the present analysis. We
note that one single ring enables us to reproduce the visibility in the
two bandheads, indicating that both the 2-0 and 3-1 bands come from the
same physical region. We also note that the CO emission region
  is inside the near infrared continuum one, the latter being
located at a radius of $R_c = 0.24_{-0.03}^{+0.06}$AU. This result corroborates 
  that the CO is emitted from a dust free region, regardless of the
  true origin of the continuum (see Sect. \ref{subsec_cont}).
At such a distance from the star, the survival of the CO gas from
photo-dissociation must however be considered. \citet{thi_1} suggested that two
mechanisms can circumvent the photo-dissociation process: the
self-shielding of the CO molecules that occurs if the column density
is high enough ($N(CO)>10^{15}\mathrm{cm}^{-2}$,
\citet{dishoeck_1}), as well as the C + OH  $\rightarrow$ CO + H
chemical reaction, which insures a continuous replenishment of this
molecule.
\subsection{The origin of the continuum near-infrared excess} \label{subsec_cont}
\begin{figure}[!t]
   \centering
   \includegraphics[width=0.4\textwidth]{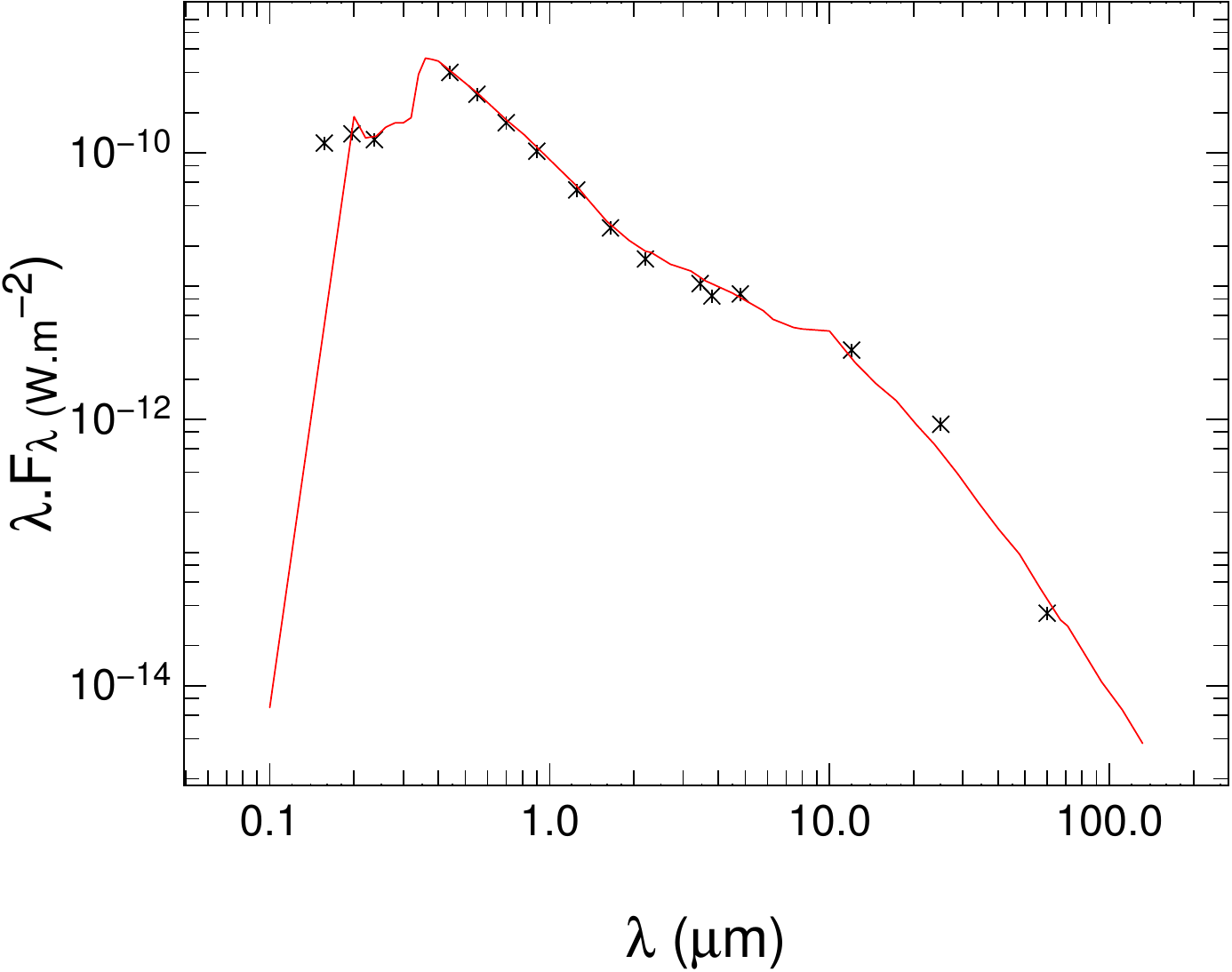}
   \caption{Spectral energy distribution of 51 Oph taken from
     literature \citep{waters_1} (crosses) and superimposed best fit
     model (solid line) using MCFOST dusty disk model. The main parameters that are fitted are the
     inner radius $R_{dust}=1.2$AU, the outer radius $R_{out}=400$AU
     scale height at 1AU $H_0 = 0.04$, the flaring exponent
     $\beta=1.1$, the grain size distribution (see text), the surface
     density exponent $p=-1.25$ and the total
     mass of dust $M_{dust} = 10^{-8}M_{\odot}$. 
     We refer to \citet{pinte_1} for a thorough
     description of the modelling.}\label{fig_results_sed_1}%
   \end{figure}
\begin{figure}[!t]
   \centering
   \includegraphics[width=0.45\textwidth]{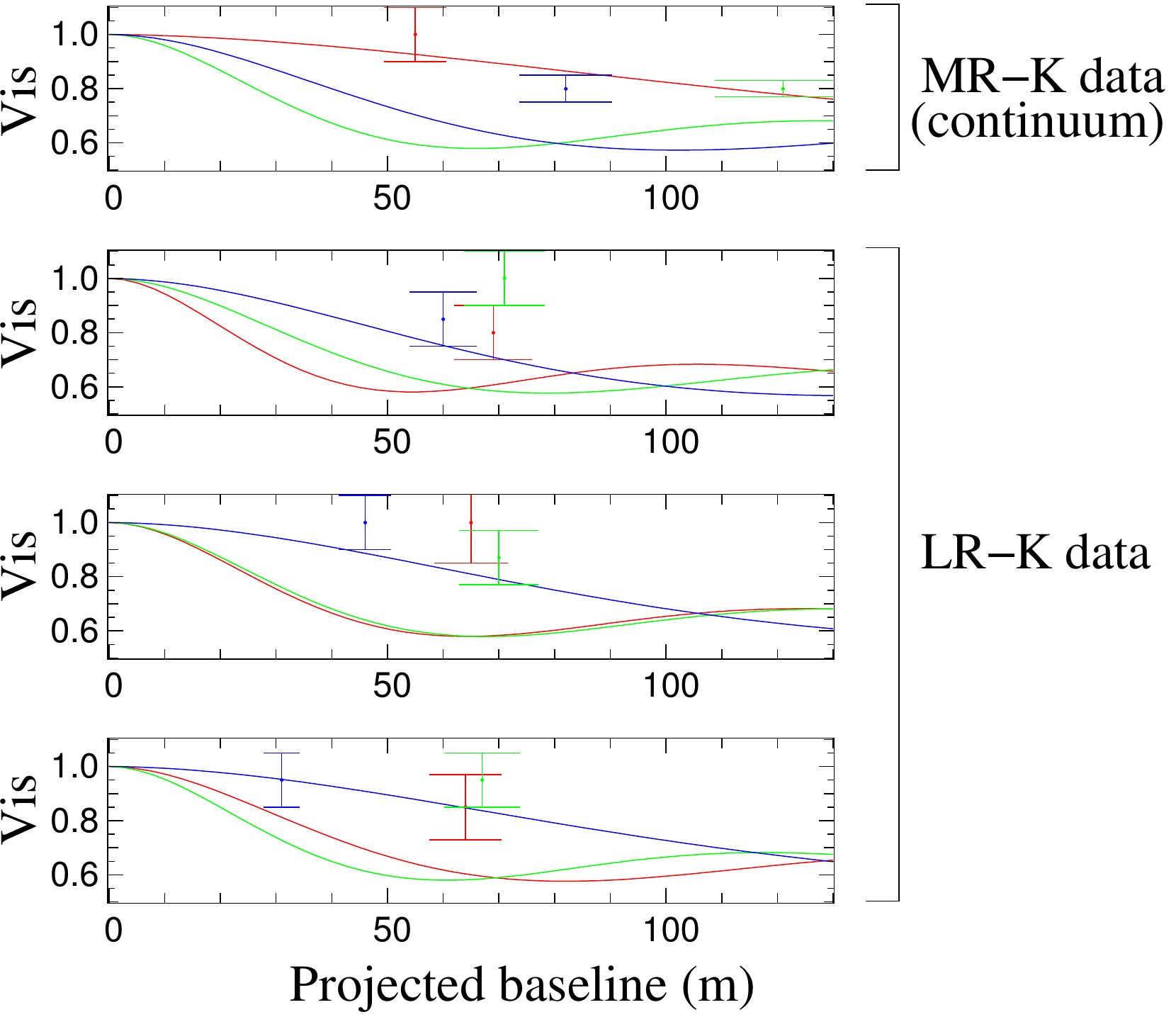}
   \caption{MR-K continuum visibilities (top panel) and LR-K
     visibilities as a function of the baseline and  and overplotted
   MCFOST model arising from SED fitting (lines). The color code is the
 same than in fig. \ref{fig_results_MRK_alone} and
 \ref{fig_results_LRK_alone}. We can see that this model, with inner
 radius equal to the dust sublimation radius, is incompatible
 with the interferometric data, with a resulting $\chi^2$ of $\sim$10.}\label{fig_results_sed_2}%
   \end{figure}
% The continuum emission is
% located at a radius of $R_c = 0.24_{-0.03}^{+0.06}$, that is 
% larger than that of the CO \textbf{bandheads}. 
The continuum excess of 51 Oph is weak and, according to the shape
of its SED \citep{waters_1},  appears to come from
a small and tenuous disk of dust which is optically thin at most
wavelengths, as already inferred by \citet{malfait_1}. 
We performed SED fitting with the MCFOST dusty disk code of \citet{pinte_1}, and found the same
results with a low amount of dust present in the disk ($M_{dust} \sim
10^{-8}M_{\odot}$). 
The result of our
modelling is shown in Fig. \ref{fig_results_sed_1}. 
This model however requires a dust
inner edge located at the dust sublimation radius, which in the case
of 51 Oph would be at $R_{dust} \sim 1.2$AU for a typical grain size
distribution of spherical particles\footnote{${\rm d}n(a) \propto
  a^{-3.7} ~{\rm d}a$, ranging from $a_{\rm min} = 0.03~\mu$m to
  $a_{\rm max} = 1~$mm, with optical constants from
  \citet{mathis_1}} and a dust sublimation temperature of $T_{dust} \sim
1500K$. Obviously this model, with such a large inner rim, is not
  compatible with our measured visibilities as shown by Fig.
  \ref{fig_results_sed_2}, and for which we calculate a $\chi^2$ of $\sim$10.  Studying the extreme case where only big grains
are present would shrink the dust sublimation radius down to $R_{dust}
\sim 0.56$AU, still at
least two times further out than the distance suggested by our data. 
%Besides, no
%satisfactory fit of the SED can be obtained from this model for small
%grains are lacking to reproduce the mid-infrared emission observed.
Interestingly, a similar problem was evidenced by
\citet{leinert_1} on the same star, the 0.5AU mid-infrared size of
51 Oph derived from their MIDI measurements being too small to be
compatible with the shape of the SED. \\ 
Several scenarios may however circumvent this apparent
    disagreement of the location of the dust evaporation radius with
    respect to our derived inner edge position. First, the gas inside the dust
sublimation may  help the
dust to survive closer in by absorbing a fraction of the stellar
radiation available to heat the dust.  According to
\citet{muzerolle_1}, the gas accreting onto the star becomes
optically thick to stellar radiation for accretion rates higher than
$\dot{M} \sim 10^{-7}M_{\odot}/yr$. The accretion rate of
$1-2.10^{-7}M_{\odot}/yr$ derived  from $\mathrm{Br}\gamma$ luminosity 
would actually place the gas in 51 Oph at the frontier between the
optically thin and optically thick regimes. \\
Another possibility comes
from the rapidly rotating nature of 51 Oph. Given its measured
rotational velocity $v\sin{i}=270$km/s \citep{dunkin_1}, and knowing that
this star is seen almost edge-on, 51 Oph would rotate at $90\%$ of its
critical velocity given by $v_{crit}=\sqrt{2GM_{\ast}/(3R_{\ast})}
\simeq 300$km/s. Therefore, 51 Oph is
likely elongated with a drop of its gravity ($g_{eff}$) from pole to
equator, and subsequently of its effective temperature \citep{zeipel_1}. 
The star being cooler at the equator than at the poles, the dust
distributed in the equatorial plane will be heated less efficiently
and the evaporation radius moved closer to the star. From the gravity
darkening law $T_{eff}^4 \propto g_{eff}$ (solid body approximation),
the effective temperature of 51 Oph at the equator can be estimated 
to be of the order of $75\%$ of that of the pole, hence lowering the sublimation
radius by a factor of 2.\\
Finally, one last explanation would be that the infrared
excess is not entirely originating from dust thermal emission.   
The gas inside the dust sublimation radius could substantially contributes 
to the near infrared energy balance. \citet{muzerolle_1} have shown
that inner optically thick gaseous zones are indeed expected to emit a
large continuum excess  in the near infrared through free-free
emission. Interestingly, this effect has been recently shown to take
place in Herbig Ae/Be stars \citep{eisner_2, isella_2, tannirkulam_1}.
\subsection{51 Oph: a classical Be star?} 
51 Oph appears to be a peculiar
source in an unusual transitional state. In the frame of Herbig Ae/Be
stars, its SED presents a near infrared excess which is too small to
account for a classical puffed-up inner rim.  In the frame of $\beta$-Pic
like stars, 51 Oph is also lacking the far infrared-excess bump
associated with the presence of an outer dusty disk, the inner disk
being emptied  by a potential forming planet \citep{malfait_1}. 
Instead these authors suggest that 51 Oph is
undergoing an alternative evolution scheme, without forming planets. 
Furthermore, the presence of strong CO overtone emission bandheads also makes 51 Oph
quite a puzzling case. Such emission requires large column
densities of warm  gas in order to produce detectable emission. Such
large column densities are rare except in sources with the largest
accretion rates \citep{najita_1}. Interestingly, the CO overtone has also been 
detected in the B9 star HD58647 \citep{berthoud_2}, for which the SED
profile is similar to that of 51 Oph \citep{malfait_1}. In HD58647 however,
there is less CO emission but its infrared excess is stronger. 
\citet{berthoud_2} thus concluded that both stars are most likely
classical Be stars surrounded by massive gaseous disk, though seen at
different evolutionary stages. This scenario is in agreement with
their high rotational velocities ($v\sin{i}=270$km/s and $118$km/s for 51
Oph and HD58647 respectively) and gives credit to the hypothesis that most of
the near infrared continuum  emission -- if not all -- is arising from the
circumstellar gas. HD58647 has been observed with the Keck
interferometer \citep{monnier_1} and these authors have derived a
rather compact size of $0.4$AU for its near infrared continuum
emission, slightly larger as that of 51 Oph for an equivalent luminosity ($L=260L_{\odot}$ for 51 Oph vs. $L=250L_{\odot}$ for HD58647).  
This is again below -- or at the very lower limit of -- the dust
sublimation radius, pointing towards the same gaseous origin for the continuum
than that of 51 Oph.
Hence, the very inner environment of this type of stars seems to follow an
intriguing evolution scheme where the dust  progressively dissipates
leaving behind a massive gas-rich, strongly accreting disk. 
As the hot dust is vanishing, the relative contribution of the gas to
the continuum infrared excess increases, its region of emission moving
closer to the star. How the circumstellar dust disappears remains rather unclear. Repeated studies combining spectroscopic detection of emission
bandheads (CO, hydrogen) and interferometric measurements in a large
sample of stars would certainly help improve our understanding on how
these disks are evolving and dissipating.
\section{Summary}
The principal results are summarized here:
\begin{itemize}
\item using the AMBER/VLTI interferometer, we have spatially and
  spectrally resolved the CO bandheads emission at $2.3\mu$m
\item the inclination of the system is constrained by our
  observations, and we confirm that the object is seen nearly edge-on. 
\item assuming simple ring models for the distribution of
    both the near infrared continuum and CO emissions, we locate the bandheads emission region at a distance of
  $0.15_{-0.04}^{0.07}$AU from the star, inside the continuum
    emission region, reinforcing the scenario in which the CO
    bandheads originate from
  the inner part of a dust free gaseous disk.
\item the continuum emission is located at a distance of
  $0.24_{-0.03}^{+0.06}$AU, at least two times closer to the star than the dust sublimation
  radius. This suggests that the inner disk of gas is playing an important role, whether by
  shielding the star light and thus allowing the dust to survive closer to
  the star, and/or by contributing to a non negligible part of the near infrared
  continuum emission. 
\end{itemize}
\begin{acknowledgements}
Authors would like to thanks C. Pinte and G. Duch\^ene for fruitful
discussions concerning the disk modelling using the MCFOST
sofware. This project was partially supported by the PRIN INAF 2006
grant "From Disks to Planetary Systems". At LAOG 
 E.T. is supported by a postdoc grant from CNRS/INSU,
France. This research is supported by {\it Agence Nationale de la
  Recherche} (ANR) of France through contract ANR-07-BLAN-0221 and by
{\it Programme National de Physique Stellaire} (PNPS) of CNRS/INSU,
France.A.N. and L.T. were  partially supported by the INAF 2005 grant
``Interferometria in-  frarossa: ottimizzazione di osservazioni
astrofisich'' and by the INAF  2006 grant ``From Disks to Planetary Systems''.  \end{acknowledgements}

\end{document}